\documentclass[twocolumn,aps,pra]{revtex4}
\usepackage{graphics,graphicx}
\usepackage{dcolumn}
\usepackage{bm}

\def\eb{\begin{equation}}   
\def\ee{\end{equation}}     
\def\ea#1{\begin{eqnarray} #1 \end{eqnarray}}   

\def\prtsq#1{{\partial^2 \over \partial {#1}^2}}
\def\ra{\rightarrow}

\def\eq#1{Eq.~(\ref{#1})}
\def\eqs#1#2{Eqs.~(\ref{#1}) and (\ref{#2})}

\def\of#1{\left(#1\right)}
\def\sof#1{\left[ {#1} \right]}

\def\shro{Schr\"odinger}
\def\H{\hat{H}}
\def\bk{\hskip -3.0pt \bigm| \hskip -3.0pt}
\def\inprod#1#2{\bigl\langle #1 \bk #2 \bigr\rangle}
\def\RE#1{#1_{\text{RE}}}
\def\IM#1{#1_{\text{IM}}}

\begin{document}


\title{On Flux Continuity and Probability Conservation in
Complexified Bohmian Mechanics}

\author{Bill Poirier}
\email{Email:  Bill.Poirier@ttu.edu}
\affiliation{Department of Chemistry and Biochemistry, and
         Department of Physics, \\
          Texas Tech University, Box 41061,
         Lubbock, Texas 79409-1061}

\begin{abstract}

Recent years have seen increased interest in complexified Bohmian
mechanical trajectory calculations for quantum systems, both as
a pedagogical and computational tool. In the latter context, it is
{\em essential} that trajectories satisfy probability
conservation, to ensure they are always guided to
where they are most needed. In this paper, probability conservation
for complexified Bohmian trajectories is considered.
The analysis relies on time-reversal symmetry
considerations, leading to a generalized expression for the
conjugation of wavefunctions of complexified variables. This in turn
enables meaningful discussion of complexified flux continuity,
which turns out {\em not} to be satisfied in general,
though a related property is found to be true. The main
conclusion, though, is that even under a weak interpretation,
probability is {\em not} conserved along complex Bohmian
trajectories.

\end{abstract}

\maketitle



\section{INTRODUCTION}
\label{intro}

Trajectory interpretations of quantum mechanics have been of
interest since the earliest days of the quantum theory. Indeed, they
even predate the \shro\ equation itself---as one finds, e.g., by
considering the Bohr-Sommerfeld quantization rule. Although the
latter was discovered to be an incorrect description of quantum
theory, it survives today in the form of the
Jeffrey-Wentzel-Kramers-Brillouin (JWKB) approximation, or more
generally, semiclassical mechanics \cite{tannor}. In this approach,
a time-evolving quantum pure state is treated as a statistical
ensemble of classical trajectories that ``carry'' approximate
quantum information, i.e. complex amplitudes. There are many
reasons, both pedagogical and practical, why semiclassical and even
classical trajectory methods may be regarded as beneficial. Surely
though, one of these must be the fact that the differential
probability, $\rho(x) dx$, is {\em conserved} along any given
trajectory---a well-known property of classical statistical
ensembles. This alone ensures that the ensemble trajectories travel
to where they are ``most needed,'' i.e. to where the probability
density is largest---a consideration that is especially important
for localized wavepacket propagation in the limit of large system
dimensionality.

Starting with Madelung in the same year as the \shro\ equation
itself \cite{madelung26} (based on matter wave ideas of deBroglie)
and evolving into a full-fledged interpretation of quantum theory
with D. Bohm and coworkers in the early 1950's
\cite{bohm52a,bohm52b}, an {\em exact} trajectory formulation of
quantum mechanics has also been developed. Over the ensuing decades,
the resultant ``Bohmian mechanical'' trajectory ensembles have been
relied upon to provide interpretational or ``analytical'' insight
into solved time-dependent quantum wavepacket propagation problems,
such as the fundamental double-slit experiment
\cite{holland,sanz00,sanz02,wangz01,yang05b}. More
recently, innovations spearheaded by members of the chemical physics
community have led to the use of quantum trajectory methods (QTMs)
as a ``synthetic'' tool, i.e. to solve the time-dependent \shro\
equation (TDSE) itself \cite{wyatt,lopreore99,mayor99}.

Though it is of great interest to compare and contrast the behavior
of quantum trajectories with their classical counterparts, we shall
do so here only as it relates to the present goals, as a detailed
discussion would take us too far afield. Most important in the
present context is the fact that Bohmian quantum trajectories {\em
also} satisfy probability conservation [\eq{pcon}]---an extremely
beneficial property for the ``synthetic'' application of QTMs, and
again, a chief reason for their utility. On the other hand, standard
Bohmian mechanical trajectories---which we shall henceforth refer to
as ``real-valued Bohmian trajectories'' for reasons that will become
clear---suffer from certain apparent drawbacks as well, some of
which can cause severe numerical difficulties for the synthetic
approach \cite{wyatt}. In particular, for non-degenerate stationary
states, all quantum trajectories are stationary fixed points---in
stark contrast to the corresponding classical trajectory orbits.

To circumvent the above problems, one approach is to follow the
semiclassical prescription of adopting a multipolar expansion of the
wavefunction, $\Psi$ \cite{poirier04bohmI,poirier07bohmalg}. This
leads to correspondence between individual (real-valued) quantum and
classical trajectories in the classical limit. A second approach,
the focus of the present paper, involves a different generalization
of Bohm's original prescription---i.e., allowing the coordinates,
$x$, and trajectory velocities, $v$, to take on complex values. The
resultant ``complex-valued Bohmian trajectories'' offer certain
advantages; for instance, they are known not to be fixed-points, in
general, for nondegenerate stationary states, so that it is possible
to achieve nontrivial trajectory dynamics in this context. Although
complex-valued Bohmian mechanics may still be in its infancy,
interest has grown tremendously in the last few years
\cite{tannor,yang05b,leacock83,tourenne86,john02,yang05a,yang06a,yang06b,yang07a,yang07b,goldfarb06,goldfarb07,goldfarb07B,goldfarb07C,chou06a,chou06b,chou07,wyatt07,sanz07,sanz07B}.
The field appears to have started in the 1980's with a paper by
Leacock and Padgett \cite{leacock83} and a less well known (and very
brief) article by Tourenne \cite{tourenne86}. More recent authors
have explored the complex Bohmian approach both for time-independent
(stationary) and time-dependent (wavepacket propagation) problems,
in both the analytical and synthetic contexts
\cite{john02,yang05a,yang06a,yang06b,yang07a,yang07b,goldfarb06,goldfarb07,goldfarb07B,goldfarb07C,chou06a,chou06b,chou07,wyatt07,sanz07,sanz07B}.

Remarkably, the all-important {\em issue of trajectory probability
conservation does not yet appear to have been addressed} in the
literature. Yet as stated previously, this is an
essential requirement if the complex-valued synthetic TDSE methods
currently under development are to have general utility for systems
larger than two or three dimensions, $d$. These methods {\em
already} require special root-finding procedures to single out the
subset of complex trajectories that happen to arrive on the
$d$-dimensional real ``axis'' $x$ at a desired final time
$t_f$---since at other times, this trajectory ensemble is described
by some non-trivial $d$-dimensional manifold embedded in the
$2d$-dimensional complex coordinate space. Alternatively, one could
in principle avoid the root search by simultaneously propagating
trajectories over the entire complex coordinate space, but
only at the expense of doubling the dimensionality of the trajectory
ensemble space, and thus enormously increasing the required number
of trajectories to compute. Such an approach would be highly
redundant from an information theory perspective, but perhaps
feasible---much like coherent state representations \cite{tannor}.

Though daunting, it may be possible to overcome the above ``$d$-doubling
problem'' \cite{sanz07} for large systems, but {\em only}
if complex quantum trajectories turn out to satisfy probability
conservation, at least approximately. This paper thus considers the
issue of probability conservation for complexified Bohmian trajectories.
The analysis ultimately relies on time-reversal symmetry considerations,
leading to a generalized definition of the conjugation operation for
functions of complexified variables, which in turn leads naturally to
a discussion of complexified flux continuity and probability conservation.
Analytical properties of complex trajectory dynamics are then considered,
and finally, several $d=1$ examples are discussed.

\section{BACKGROUND}
\label{background}

Real-valued Bohmian mechanics begins with the Madelung-Bohm ansatz
for the wavefunction \cite{madelung26,bohm52a,bohm52b},
\eb
     \Psi(x,t) = R(x,t) e^{i S(x,t)/\hbar}. \label{oneLMB}
\ee
The decomposition above is essentially unique, by virtue of the fact
that $R$ and $S$ are taken to be real-valued, and $R>0$. By substituting
\eq{oneLMB} into the TDSE, and gathering real and imaginary terms
separately, one obtains the following two real-valued equations:
\ea{
{\partial R \over \partial t} & = & {-1 \over 2m} \of{2 R' S' + R S''}
     \label{Rdotuni} \\
{\partial S \over \partial t} & = & - \sof{ {{S'}^2\over 2m} + V  -
          {\hbar^2 \over 2m} {R'' \over R} }, \label{Sdotuni} }
where the primes denote spatial differentiation.
Equation~(\ref{Sdotuni}) is the ``quantum Hamilton-Jacobi equation''
\cite{bohm52a}, which imparts a classical-field-theory-like \cite{goldstein}
interpretation to quantum wavepacket propagation, provided that:
(a) $S'(x,t)$ is interpreted as trajectory momentum;
(b) an additional ``quantum potential,''
$Q(x,t)=-(\hbar^2/2m) \of{R''/ R}$, is added to the true potential,
$V(x)$, to determine the trajectory dynamics.

Equation~(\ref{Rdotuni}) is the flux continuity equation, which under
the above interpretations, is identical to trajectory probability
conservation. In particular, \eq{Rdotuni} can be rewritten as
\ea{ \partial \rho(x,t) \over \partial t & = & - j'(x,t),
                              \quad\text{where} \label{cont} \\
     \rho(x,t) & = & R^2(x,t) \quad\text{is probability density,}  \\
     j(x,t) & = & \rho(x,t) v(x,t) \quad\text{is probability flux, and}
                                           \label{fluxreal} \\
     v(x,t) & = & S'(x,t)/m \quad\text{is trajectory velocity.} \label{velo} }
Note that the flux $j(x,t)$ above accords with the usual quantum definition,
i.e.,
\eb
     j = {\hbar \over 2 i m} \of{\Psi^* \Psi' - {\Psi'}^* \Psi}. \label{frp}
\ee
Note also that the flux is independent of potential energy as is reasonable,
i.e. $V(x)$ enters {\em only} into the dynamical equation for trajectory
evolution, \eq{Sdotuni}. In any case, \eq{cont} implies conservation
of differential probability along a trajectory, i.e.
\eb
     {d[\rho(x,t) dx] \over dt} = 0, \label{pcon}
\ee where $d/dt$ refers to the total (hydrodynamic) time derivative.
For simplicity, the above equations have been written as if the
dimensionality is $d=1$, but they are meant to refer also to the
multidimensional case, as the $d>1$ generalizations are
straightforward.

In complex Bohmian mechanics, an altogether different ansatz is employed:
\eb
     \Psi(x,t) = e^{i S(x,t)/\hbar}. \label{compans}
\ee
In effect, the real-valued amplitude $R$ of the \eq{oneLMB} decomposition
is ``subsumed'' into the exponent to form the imaginary part of the
now-{\em complex}-valued action, $S$. Substitution into the TDSE now yields the
{\em single}, complex-valued equation,
\eb
     {\partial S\over \partial t}  =  - \sof{ {{S'}^2\over 2m} + V  -
          {i \hbar \over 2m} S'' }, \label{Sdot}
\ee
which can be interpreted as a complex quantum Hamilton Jacobi
equation. Equation~(\ref{Sdot}) is the starting point for the quantum
Hamilton-Jacobi formalism, regarded as one of the nine fundamental
formulations of quantum mechanics.\cite{leacock83,yang05a,styer02}

Unlike the real-valued \eq{Sdotuni}, \eq{Sdot} contains
{\em all} of the information present in $\Psi$, leading some authors
to conclude that the complex version is the more fundamental
\cite{yang05a}. On the other hand, in going from real- to
complex-valued formulations we have apparently lost the flux
continuity relation altogether! Thus, if there is indeed a
probability conservation property for complex trajectories, it is
neither manifest in, nor independent from, \eq{Sdot}. Note that in
certain contexts, it is possible to extract a complex
energy-momentum conservation relation from \eq{Sdot}
\cite{yang05a,baker-jarvis03}. However, this is not directly useful
for synthetic applications, vis-a-vis the guidance of trajectories
to high-probability regions where they are needed most.

\section{COMPLEXIFICATION: TIME REVERSAL SYMMETRY AND
GENERALIZED COMPLEX CONJUGATION}
\label{complexification}

To make progress with regard to probability conservation, it seems
clear that \eq{Sdot} must be written in a different, but
mathematically equivalent, form. In comparing \eq{oneLMB} to
\eq{compans}, which emphasizes the action $S$ over amplitude $R$, it
seems clear that the opposite procedure should be applied to
emphasize the latter---i.e. the real-valued $S$ of \eq{oneLMB}
should be ``pulled down'' from the exponent to form a complex-valued
amplitude, $R$. But this would simply yield $\Psi$ itself as the
appropriate quantity to work with. Thus, the TDSE in its standard
form (i.e. in terms of $\Psi$) should be a reasonable starting point
for analyzing flux continuity---albeit a {\em complexified} version,
with $x \in {\cal R}$ replaced with $z \in {\cal C}$.
Complexification of the TDSE per se offers no inherent difficulty
provided that $V(x)$ and the initial wavepacket, $\Psi(x,t=0)$, are
both analytic functions, as everything can then be uniquely lifted
from the real $x$ axis to the complex $z$ plane in accord with the
usual rules of analytic continuation \cite{arfken}. On the other hand,
establishing a complexified probability density, $\rho(z)$, does
pose a bit of a problem, as the traditional definition $\rho(x) =
\overline{\Psi(x)} \Psi(x)$ in terms of the conjugate function
$\overline{\Psi}(x) = \Psi^*(x)= [\Psi(x)]^*$ is in general {\em
not} analytic when $x$ is replaced with $z$.

Two straightforward candidates for a complexified $\rho(z)$ quantity
are presented below:
\begin{enumerate}
\item{Define $\rho(z) = \Psi^*(z) \Psi(z)$.}
\item{Define $\rho(z)$ as the analytic continuation of $\rho(x)$.}
\end{enumerate}
Option 1. has the presumed advantage that $\rho(z)$ is positive and
real-valued everywhere in the complex plane, but suffers from the
severe drawback that $\rho(z)$ is not an analytic function. This
approach is being considered by other authors \cite{sanz07B}.
Option 2. offers analyticity, but yields complex-valued probability
densities off of the real axis, and bypasses the more fundamental
issue of non-analytic conjugate wavefunctions, which must still be
resolved. Analyticity is an enormous advantage, for it means that
familiar expressions such as $\int_{-\infty}^{+\infty} \rho(z) dz =
1$ represent true contour integrations with path-independent
meaning. Thus, integration contours may be deformed away from the
real axis---which is especially important for synthetic TDSE
applications, given that the contour is essentially the
(time-evolving) trajectory ensemble manifold.  In any case, complex
probability values are likely unavoidable in complexified space, for
even if the density $\rho(z)$ is real-valued, the differential
probability itself, $\rho(z) dz$, need not be.

We will resolve the matter by directly addressing the more general
and fundamental issue of wavefunction conjugation on complexified
spaces. One approach to this problem is to invoke
charge/parity/time-reversal (CPT) symmetry \cite{sakurai}---an idea
that was introduced previously in the specific context of
non-Hermitian Hamiltonian operators
\cite{yang05a,bender02,swanson04}. Similar ideas can be applied in
the present case of complexified Hermitian Hamiltonians, although
for general potentials, only time-reversal symmetry is relevant. Let
$\Psi(z,t)$ be a solution of the complexified TDSE ($z \in {\cal
C}$, but $t \in {\cal R}$):
\eb
          i \hbar {\partial \Psi(z,t) \over \partial t} =
          -{\hbar^2 \over 2m} {\partial^2 \Psi(z,t) \over \partial z^2}
          + V(z) \Psi(z,t)  \label{complexTDSE}
\ee
Complex conjugating both sides and applying explicit time-reversal, i.e.
$t \ra -t$, yields
\ea{
          i \hbar {\partial \sof{\Psi(z,-t)}^* \over \partial t} & = &
          -{\hbar^2 \over 2m} \sof{{\partial^2 \Psi(z,-t) \over \partial z^2}}^*\nonumber\\
                    & & + [V(z)]^* \sof{\Psi(z,-t)}^*.
          \label{usualTR}}

The above is the usual means of deriving the effect of the
antiunitary time-reversal operator on a wavefunction, for
real-valued coordinates \cite{sakurai}. For the complexified case,
however, it is clear that \eq{complexTDSE} is {\em not} equivalent
to \eq{usualTR}, because $[V(z)]^* = V^*(z)$ is not equivalent to
$V(z)$ off of the real axis. Instead, we must introduce the
additional and final step of replacing the coordinate $z$ with its
complex conjugate, $z^*$. From the Schwartz reflection principle
\cite{arfken}, $V^*(z^*) = V(z)$, because $V(x)$ is real-valued.
Thus,
\eb
          i \hbar {\partial \Psi^*(z^*,-t) \over \partial t} =
          -{\hbar^2 \over 2m} {\partial^2 \Psi^*(z^*,-t)
          \over \partial z^2} + V(z) \Psi^*(z^*,-t),
          \label{complexTR}
\ee so $\Psi^*(z^*,-t)$ is also a solution of the TDSE, the proper
manifestation of time-reversal symmetry on the complexified space.
Note that the final step above is also consistent with the approach
of H\"uber, Heller, and Littlejohn \cite{huber88}, which may be
regarded as a semiclassical approximation to complexified Bohmian
Mechanics. In fact, there is a very close connection between
the present approach and complex semiclassical mechanics, with the
latter emerging as the first-order truncation of an infinite series
expansion of the former, as elucidated by Tannor and coworkers
\cite{goldfarb06,goldfarb07,goldfarb07B,goldfarb07C}.

Following Bender's approach \cite{bender02,swanson04}, the
corresponding conjugation operation is therefore
\eb
     \overline{f(z)} = f^*(z^*) = g(z). \label{conj}
\ee Note that along the real axis, i.e. $z=x$, \eq{conj} is
equivalent to the standard conjugation operation. Most importantly
however,  we find that $f^*(z^*)$ is an {\em analytic function} in
the original variable $z$, in that it satisfies the Cauchy-Reimann
conditions \cite{arfken}---unlike, say, $f(z^*)$ or $f^*(z)$.
Equation~(\ref{conj}) also satisfies the mathematical definition of
a conjugate linear map \cite{arfken}, unlike $f(z^*)$ or $f^*(z)$.
Note that $f^*(z^*)$, is {\em not} equivalent to $f(z)$ in general.
We will therefore sometimes refer to $f^*(z^*)$ as ``$g(z)$,'' to
emphasize both its distinctness from $f(z)$ and also its analyticity.
The latter property can be demonstrated explicitly from a Taylor
expansion:
\eb
     f(z) = \sum_{k=0}^\infty C_k z^k \quad; \quad
     g(z) = \sum_{k=0}^\infty C_k^* z^k.
\ee

Having shown that $\overline{f(z)}$ is analytic, it follows trivially
that the complexified inner product integration,
\eb
     \inprod{f}{h} = \int_{-\infty}^{+\infty} g(z) h(z) dz,
\ee
is contour path-independent, as the integrand itself is analytic
[assuming analytic $h(z)$]. In particular, this implies that
\eb
     \rho(z) = \overline{\Psi(z)} \Psi(z) \label{rhodef}
\ee
is analytic, and moreover, is equivalent to option 2. described above.

Note that both $f(z) = V(z)$ and $f(z) = \rho(z)$ share the special
property that $\overline{f(z)} = g(z) = f(z)$. This is nothing but
the Schwartz reflection principle, and is true whenever $f(z)$ is
analytic and $f(x)$ is real. We find it convenient to refer to such
a function as a REAL function---even though it is understood that
$f(z)$ is not real-valued off of the real axis. Similarly, an IMAG
analytic function $f(z)$ is defined such that $\overline{f(z)} =
g(z) = -f(z)$, and is pure imaginary-valued along the real axis, but
not necessarily elsewhere. Any analytic function $f(z)$ can be
decomposed into a sum of REAL and IMAG parts, which we demonstrate
via explicit construction: \ea{
     f(z) & = & \RE{f}(z) + \IM{f}(z), \quad \text{where} \label{REIM} \\
     \RE{f}(z) & = & \sof{{f(z) + f^*(z^*)} \over 2} \nonumber \\
     \IM{f}(z) & = & \sof{{f(z) - f^*(z^*)} \over 2} \nonumber
}
Note that $\RE{f}(z)$ and $\IM{f}(z)$ are themselves analytic---unlike say,
$\text{Re}[f(z)]$ and $\text{Im}[f(z)]$ (although $\RE{f}(x) = \text{Re}[f(x)]$,
etc.) Note also that the analytic derivative $f'(z) = d f(z)/dz$ of a REAL(IMAG)
function $f(z)$ is also REAL(IMAG). Finally, the product of a REAL and REAL
(or IMAG and IMAG) pair of functions is REAL, whereas the product of REAL and
IMAG functions is IMAG.

\section{FLUX CONTINUITY AND PROBABILITY CONSERVATION}
\label{fcpc}

Our next goal is to define a complexified flux, from which to derive
a corresponding flux continuity relation. We start with the
velocity field $v(z,t)$, which from \eqs{velo}{compans} is given by
\cite{leacock83,john02,yang05a,goldfarb06}
\eb
     v(z,t) = - {i \hbar \over m} {\Psi'(z,t) \over \Psi(z,t)}.
     \label{vee}
\ee In general, $v(z,t)$ is neither REAL nor IMAG. Along the real
axis, $\RE{v}(x,t)$ is the ``flow velocity'' (i.e. the standard
velocity field of real-valued Bohmian mechanics, closely related to
hydrodynamics), whereas
$\IM{v}(x,t)$ is known as the ``Einstein osmotic velocity,''
associated with wavepacket spreading, or diffusion in stochastic
quantum mechanics \cite{wyatt,nelson66,guerra81,pavon95}. However,
$\IM{v}(x,t)$ is not used in the stochastic context to generate
trajectories per se, or otherwise venture off of the real coordinate
axis. Hirschfelder did use the imaginary velocity to create
real-valued trajectories or ``streamlines,'' but only along the real
axis, by replacing $t$ with $i t$ \cite{hirschfelder74}. In
complexified Bohmian Mechanics, the imaginary velocity is directly
responsible for transporting the complex trajectories off of the
real axis. Of key significance for the present approach is
that {\em both velocity components can be given significance off of
the real axis}, due to the \eq{REIM} decomposition, and the fact
that $\RE{v}(z,t)$ and $\IM{v}(z,t)$ are analytic functions,
provided $v(z,t)$ is (at least locally) analytic. Note that global
analyticity of $\Psi(z,t)$ does not necessarily imply the same
property for $v(z,t)$ (Sec.~\ref{analytic}).

The complexified flux is naturally defined from \eqs{rhodef}{vee},
and the analytic continuation of \eq{fluxreal}, as
\eb
     j(z,t) = v(z,t) \rho(z,t) = - {i \hbar \over m} \Psi^*(z^*) \Psi'(z),
     \label{flux}
\ee
which based on the multiplication rules given in Sec.~\ref{complexification},
is again neither REAL
nor IMAG. Note that the conjugation operation of \eq{conj} commutes
with spatial differentiation, i.e. $[d f(z^*)/d z^*]^* = g'(z)$, where
$g(z) = f^*(z^*)$. Thus, the meaning of expressions such as the following
are unambiguous:
\ea{
     \RE{j}(z,t) & = &\RE{v}(z,t) \rho(z,t)\nonumber\\
     & = &
     - {i \hbar \over 2 m} \sof{\Psi^*(z^*) \Psi'(z) -  \Psi(z){\Psi'}^*(z^*)}
     \label{fluxRE} \\
     \IM{j}(z,t) & = &\IM{v}(z,t) \rho(z,t)\nonumber\\
     & = &
     - {i \hbar \over 2 m} \sof{\Psi^*(z^*) \Psi'(z) +  \Psi(z){\Psi'}^*(z^*)}
     \label{fluxIM}
}
Note that along the real axis, $\RE{j}(x,t)$ is equivalent to the usual
real-valued quantum flux of \eq{frp}. The imaginary flux, $\IM{j}(z,t)$,
in contrast, does not appear to have been considered previously
in the literature, even when restricted to the real coordinate axis.
Off of the real axis, $j$ and $v$ point in different directions.

A flux continuity relation should presumably involve the divergence of the flux.
From \eqs{fluxRE}{fluxIM} the REAL and IMAG components are found to be:
\ea{
     \RE{j}'(z,t) & = &
     - {i \hbar \over 2 m} \sof{\Psi^*(z^*) \Psi''(z) -  \Psi(z){\Psi''}^*(z^*)}
     \label{divfluxRE} \\
     \IM{j}'(z,t) & = &
     - {i \hbar \over 2 m} \big[     \Psi^*(z^*) \Psi''(z) +  \Psi(z){\Psi''}^*(z^*)\nonumber\\
                               & & \qquad\qquad  + 2 \Psi'^*(z^*)\Psi'(z) \big] \label{divfluxIM}}
To relate the above to the time derivative of $\rho(z,t)$ as defined in
\eq{rhodef} requires the time-derivative of $\overline{\Psi(z,t)}$. By
replacing $t\ra -t$ in \eq{complexTR}, this is found to be
\eb
          i \hbar {\partial \overline{\Psi(z,t)} \over \partial t} =
          +{\hbar^2 \over 2m} \overline{\Psi(z,t)}''
            - V(z) \overline{\Psi(z,t)}
          \label{complexTRTR}
\ee
Multiplying \eq{complexTDSE} by $-(i/\hbar) \overline{\Psi(z,t)}$, and
adding to $-(i/\hbar) \Psi(z,t)$ times \eq{complexTRTR} then yields the
complexified flux continuity relation
\eb
     {\partial \rho(z,t) \over \partial t} = - \RE{j}'(z,t). \label{compcont}
\ee
The above procedure is similar to that used to derive the real-valued
flux continuity equation; however, it only works here by virtue of the
Schwartz reflection principle. Indeed, a continuity relation based on
{\em any} definition of $\overline{\Psi(z,t)}$ other than \eq{conj} would
result in a {\em flux quantity that depends on the potential energy}---a highly
unphysical scenario that we reject out of hand.

On the other hand, the continuity relation of \eq{compcont} involves only
the REAL component of flux, rather than $j(z,t)$ itself. This makes
perfect sense when one considers that $\rho$ itself is REAL, and therefore
its time derivative must also be REAL. However, this has nontrivial ramifications
vis-a-vis probability conservation for complex trajectories generated from
$v(z,t)$ rather than from say, $\RE{v}(z,t)$.  We will have more to say on
this topic in a moment, but first we point out some additional noteworthy
aspects of \eq{compcont}. In particular,
\eb
     \int_{-\infty}^{+\infty} {\partial \rho(z,t) \over \partial z} dz = 0,
\ee
i.e. the total probability conservation property, which has a contour
path-independent meaning; it is in any event clear that integration of
\eq{divfluxRE} along the real axis is zero. In the specific case of nondegenerate
stationary states, one can further show that
$ \partial \rho(z,t) / \partial t = 0$ everywhere. Thus, even though $\rho$ has
a nontrivial complex phase off of the real axis, this phase does not evolve over
time, so that $\rho(z,t)$ is truly stationary everywhere. Note that without
loss of generality, $\Psi(z,t) = \psi(z) e^{- i \omega t}$ with $\psi(z)$
REAL in this case, implying that
\eb
\rho(z,t) = \sof{\psi(z)}^2. \label{rhorel}
\ee

Since the continuity relation in \eq{compcont} is given in terms of
$\RE{j}(z,t)= \RE{v}(z,t) \rho(z,t)$, it immediately follows via
analytic continuation that $d [\rho(z,t) dz] / dt = 0$, provided the
complex trajectories are obtained from $\RE{v}(z,t)$ rather than
$v(z,t)$. Along the real axis, the $\RE{v}$ trajectories are just
the standard real-valued Bohmian trajectories, so the orbit is the
real axis itself. Off of the real axis, the $\RE{v}(z,t)$ are quite
nontrivial, and might in principle be considered for dynamical
purposes, with a ready-built trajectory probability conservation
property. There are at least two drawbacks to this arrangement
however: (1) $\RE{v}$ trajectories that start off of the real axis
never intersect the real axis; (2) nondegenerate stationary states
still have $\RE{v}(z,t)=0$ everywhere in the complex plane. In any
event, all of the previous literature on complexified Bohmian
mechanics uses $v(z,t)$ rather than $\RE{v}(z,t)$ trajectories. Note
that for nondegenerate stationary states, $v(z,t) = \IM{v}(z,t)$, so
that {\em all} of the trajectory dynamics is due to the imaginary
velocity in this case. Again, because $\IM{v}(z,t)$ is not pure
imaginary-valued, this does {\em not} imply trivial ``vertical''
(parallel to imaginary axis) trajectory orbits in the complex plane.
In general, stationary state orbits are vertical only where they
intersect the real axis, about which the orbits display reflection
symmetry; otherwise, they are quite arbitrary, often recrossing
the real axis again at a different point (Sec.~\ref{results}).
The ground state of the harmonic oscillator system, for instance,
is characterized by concentric circular
orbits \cite{wyatt,john02,yang05a,yang06b}.

For the more general case of nonstationary wavepacket propagation,
it is easily shown that $d [\rho(z,t) dz] / dt \ne 0$ in general,
for trajectories obtained from $v(z,t)$. Thus, in a literal sense,
probability conservation along trajectories is {\em not} satisfied.
On the other hand, it might be argued that such a strong form of
probability conservation should not be required. In particular, this
relates to the fundamental question of how physical observables are
to be interpreted in complexified space, and how these relate to
real measurements. The impression one derives from the literature
is that observables are to have physical meaning {\em only} when
evaluated along the real axis. On the other hand, a theory such
as that presented here, which allows all quantities to be analytically
continued, may shed additional light into this question, in that
knowledge of said quantities along an essentially arbitrary
$d$-dimensional manifold implies knowledge along the real axis.

Returning to the issue of probability conservation, if we presume
that $\rho(z,t) dz$ has no direct physical meaning off of the real
axis, then it may not be appropriate to impose \eq{pcon} throughout
the entire complex plane. Instead, let us imagine a particular $v(z,t)$
trajectory, $z(t)$, which crosses the real axis at time $t_i$, and
again at a later time $t_f$, i.e. $z(t_i) = x_i$ and $z(t_f) = x_f$
(Fig.~\ref{mapfig}).
We stipulate on physical grounds that it should be sufficient to
satisfy the following {\em weak} probability conservation condition:
\eb
     \rho(x_i,t_i) dx_i = \rho(x_f,t_f) dx_f \label{pconweak}
\ee
However, even \eq{pconweak} turns out not to be satisfied
in general---even under the most favorable situation where $\Psi$ is
a non-degenerate stationary state and $v(z)$ is analytic.

\section{ANALYTIC PROPERTIES AND TRAJECTORY DYNAMICS}
\label{analytic}

\begin{figure}
\includegraphics[scale=0.8]{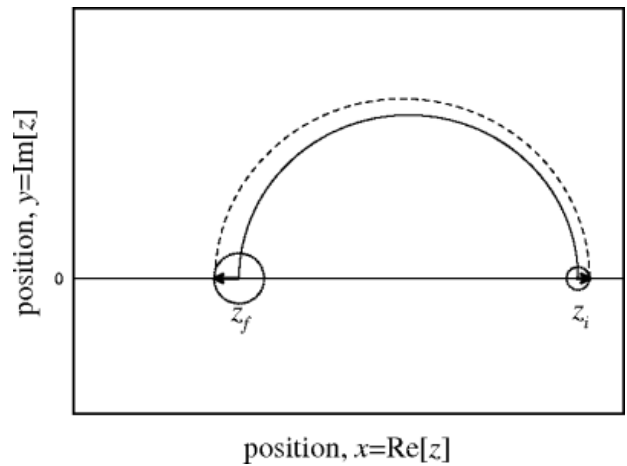}
        \caption{Two closed complex trajectory orbits for a nondegenerate
                 stationary state over half a period, $\Delta t = T/2$. All units
                 are atomic units. Solid
                 curve denotes main trajectory, starting at $(z_i=x_i,t_i)$ and
                 ending at $(z_f=x_f,t_f)$. Dashed curve denotes neighboring
                 trajectory, starting at $(x_i+dx_i)$. The function
                 $f^{(t_i,t_f)}(z)$ maps $z_i$ to $z_f$, and the small
                 differential circular disk $z_i + dz_i$ on the right to the small
                 differential circular disk $z_f + dz_f$ on the left. Right and left
                 arrows denote (directional) $dx_i$ and $dx_f$ intervals,
                 respectively; note the change in sign.}
        \label{mapfig}
\end{figure}

To demonstrate that \eq{pconweak} is not satisfied, it is helpful to
introduce the map $f^{(t_i,t_f)}(z_i) = z_f$, corresponding to the
trajectory that connects the initial point $(z_i,t_i)$ to the final
point $(z_f,t_f)$ (Fig.~\ref{mapfig}).
This in turn requires the trajectory guidance equation,
\eb
     {dx \over dt} = v(x,t) = - {i \hbar \over m} {\Psi'(z,t) \over \Psi(z,t)}.
\ee
Regarded as a function of $z_i$, $f^{(t_i,t_f)}(z_i)$ is
essentially a continuous composition of the generator map $v(z,t)$.
Physically, it represents the effect of time evolution on the system,
from the initial time $t_i$ to the final time $t_f$, and is standard
in classical field and semiclassical theories \cite{tannor,goldstein,arnold}.
Note that the time-dependence of $\Psi(z,t)$ is analogous to classical
dynamics under a time-dependent potential, thus requiring
{\em both} superscript time parameters, $t_i$ and $t_f$, rather than just
the difference, $\Delta t = (t_f - t_i)$.

If $v(z,t)$ is presumed to be analytic, at least locally in the
vicinity of some given trajectory, then $f^{(t_i,t_f)}(z_i)$ will
also be locally analytic. {\em Globally} however, $f^{(t_i,t_f)}(z_i)$ is
usually {\em not} analytic even if $v(z,t)$ itself is globally analytic,
due to branch cuts that can arise in the function $f^{(t_i,t_f)}(z_i)$.
In most cases, e.g. when $\Psi(x,t)$ has nodes, $v(z,t)$ is not (globally)
analytic, but ``meromorphic,'' meaning that it is globally analytic apart
from a discrete set of simple poles \cite{arfken}. In this case also,
$f^{(t_i,t_f)}(z_i)$ is usually not meromorphic itself (though it must
have simple poles), but also exhibits branch cuts. The branch cuts
delineate regions of the $z_i$ domain that get mapped discontinuously
under $f^{(t_i,t_f)}(z_i)$ to different regions of the range. In the trajectory
interpretation, two initially nearby trajectories, starting differentially close
to each other at $t_i$ but on either side of a branch cut, wind up far apart
at $t_f$. The branch cuts themselves must therefore correspond to
``separatrix''-type trajectory orbits (Fig.~\ref{trajfig}).

On the other hand, for trajectories sufficiently far from a separatrix
or simple pole, $f^{(t_i,t_f)}(z_i)$ may be regarded as locally analytic.
Consequently, points $z = z_i+dz_i$ lying within a
differentially small circle centered at $z_i$ get mapped under $f^{(t_i,t_f)}(z)$
to another differentially small disk of points
$f^{(t_i,t_f)}(z_i+dz_i) = z_f + dz_f$, as in Fig.~\ref{mapfig}.
Due to local analyticity \cite{arfken}, the ratio of differentials, i.e.
$dz_f/dz_i$, is independent of the magnitude or direction of $dz_i$:
\eb
     {dz_f \over dz_i} = {f^{(t_i,t_f)}}' (z_i)
\ee
Thus, the spatial derivative function, ${f^{(t_i,t_f)}}'(z_i)$ specifies how the
differential volume element, $dz$, is transformed via trajectory evolution over
the specified time interval. For a given initial point, $(z_i,t_i)$,
we can, by varying $t_f=t$, obtain a transformed $dz$ for every
point along the $z(t)$ trajectory. In particular,
\eb
     {dz \over dz_i} =  {f^{(t_i,t)}}' (z_i) = r^{(z_i,t_i)}[z(t)],
     \label{rres}
\ee
where it is understood that the trajectory, $z(t)$, passes through
$(z_i,t_i)$. Note the new quantity, $r^{(z_i,t_i)}[z(t)]$,
which---though equivalent to ${f^{(t_i,t)}}' (z_i)$---is introduced
so as to be regarded as a function of the {\em final} trajectory
point $z(t)$ (at arbitrary time $t$), rather than the initial point
$z_i$. Note also that $r^{(z_i,t_i)}(z_i) = 1$.

For differentially small time increments, $\Delta t = dt$, it can easily
be shown using standard statistical or hydrodynamical arguments that
$dz_f = \sof{1 + v'(z_i,t_i) dt} dz_i$ \cite{wyatt}, implying that
\eb
     {d r^{(z_i,t_i)} [z(t)] \over dt} = v'[z(t),t] \, r^{(z_i,t_i)} [z(t)]
     \label{rdot}
\ee
The solution for $r^{(z_i,t_i)}[z(t)]$ is therefore found to be
\eb
     r^{(z_i,t_i)}[z(t)] = \exp{\int_{t_i}^{t} v'[z(t'),t'] dt'}, \label{rsol}
\ee
where the integration is along the trajectory.

The general global solution to \eq{rsol} above is not
straightforward; in particular, it is not apparent how to obtain
$r^{(z_i,t_i)}$ values for $(z,t)$ points off of the $z(t)$
trajectory. However, this is easily achieved for the specific case
of nondegenerate stationary states, on which we will focus for the
remainder of this paper. In this context, $v(z,t) = v(z)$ is
independent of (final) time, and similarly, $r^{(z_i,t_i)} = r^{z_i}$ is
independent of initial time, as appropriate for the resultant
``time-independent potential'' dynamics. A simple change of variables in
\eq{rsol} above then leads to
\eb
     r^{z_i}(z) = {v(z) \over v(z_i)}. \label{plug}
\ee
Since $v(z) = \IM{v}(z)$ is pure IMAG, for trajectories
initiating on the real axis, $z_i = x_i$, the function $r^{x_i}(z)$
is pure REAL. Thus, whenever the trajectory $z(t)$ recrosses the
real axis, at $(z_f=x_f,t_f)$, the initial real-valued volume
element $dz_i = dx_i$ is transformed to a final volume element $dz_f
= dx_f$ that is {\em also} real-valued (Fig.~\ref{mapfig})---a
nontrivial but essential prerequisite for any \eq{pconweak}-type
relation, since $\rho(x_i,t_i)$ and $\rho(x_f,t_f)$ are also be real-valued.
In particular, we obtain
\eb
     {dx_i \over \IM{v}(x_i)} = {dx_f \over \IM{v}(x_f)} \label{dxrel}
\ee
However, it is not necessary to restrict oneself to $x_f$ values that are
connected to $x_i$ via trajectories; in fact, the above relation is true for
completely general values of $x_i$ and $x_f$.

Substituting \eq{vee} into \eq{dxrel}, and exploiting \eq{rhorel},
we derive the following relation:
\eb
     {\rho(x_i) \over \rho'(x_i)}  dx_i =
     {\rho(x_f) \over \rho'(x_f)}  dx_f. \label{newweak}
\ee
Equation~(\ref{newweak}) above is similar to, but clearly
distinct from, the desired probability conservation condition,
\eq{pconweak}. The latter is thus satisfied if and only if
$|\rho'(x_i)|=|\rho'(x_f)|$, which---though not true in general---is
true for the special case of symmetric analytic velocity fields
$[v(-z) = -v(z)]$ with a single stationary point at the origin.

Note that for $f^{\Delta_t}(z_i)$ now depends only the time
interval, $\Delta_t$. It is of interest to relate $v(z)$ to
$f^{\Delta_t}(z_i)$ via $r^{x_i}(z_f)$. With $z=z_i$ and $f =
f^{\Delta_t}(z_i) = z_f$, one obtains
\ea{
     {df \over dz} & = & r^{x_i}(z_f)  =  {v(f) \over v(z)}, \\
     w(f) & = & w(z) + \text{const},
}
where $w(y)$ is the integral $\int dy/v(y)$. Clearly, $w(y)$, and thus
$f^{\Delta_t}(z)$, will not in general be globally analytic, even if $v(z)$
itself is.

The role of stationary points, i.e. $z_0$ such that $v(z_0)=0$, is
important in trajectory dynamics. These occur where $\psi'(z)=0$,
i.e. the non-node zeroes of $\rho'(z) = 0$. When $\psi(z)$ has
nodes, it is easy to demonstrate that $v(z)$ has a simple pole at
each node, and is therefore meromorphic at best. Conversely,
node-free analytic wavefunctions $\psi(z)$ can often lead to
globally analytic velocity fields. In the vicinity of nodes, $v(z)$
directs initially neighboring trajectories to very different final
destinations, thus leading to separatrix-like trajectories and
branch cuts in $f^{\Delta_t}(z)$ (Fig.~\ref{trajfig}). However,
unlike separatrix orbits in classical phase space, $v(z)$ does not
approach zero as the node is approached---because the poles are not
themselves stationary points. Note that separatrix-like trajectories,
and associated $f^{\Delta_t}(z)$ branch cuts, can arise even when
$v(z)$ is globally analytic (Sec.~\ref{results}).

In the neighborhood of stationary points, one can apply the standard
velocity field linearization method \cite{arnold} to determine the
behavior of neighboring trajectories.
In particular, $v(z)\approx A_1(z-z_0)$, where
$A_1$ is a complex constant . If $A_1$ is pure imaginary, then the
neighboring trajectories are circular and closed, with frequency $\omega
= |A_1|$ and period $T=2 \pi/|A_1|$. This is always the case for
stationary points on the real axis, i.e. $z_0=x_0$, which we now
consider in greater detail. Note that due to reflection symmetry
about the real axis, the first crossing of the real axis occurs at
time $T/2$, at location $x_f = x_0 - (x_i-x_0)$, for differentially
small $(x_i-x_0)$. Thus, in some small neighborhood of $x_0$, the
map $f^{T/2}(x)$ is real-valued, which in turn implies that
$f^{T/2}(z)$ is REAL over the domain around $x_0$ for which
$f^{T/2}(z)$ is analytic. This in turn implies that {\em all
trajectories within the domain of analyticity have period $T$}, even
well away from the neighborhood where the velocity field is linear
and the trajectory orbits circular.

For stationary points off of the real axis, it may still be true
that $A_1$ is pure imaginary, leading to closed neighboring
trajectories, and essentially ``conservative'' dynamics. However, $A_1$
may also be complex-valued, resulting in aperiodic trajectories that
spiral in or out, with the stationary point respectively an
attractor or repeller (with the former akin to dissipative
dynamics). Such stationary points $z_0$ come in complex conjugate
pairs, with the $A_1$ of one the complex conjugate of the other,
thus implying that one point of the pair is an attractor and the
other a repeller. It is possible for trajectories to flow directly
from an attractor to its conjugate repeller, crossing the real axis
in a region without closed orbits.

\section{SPECIFIC EXAMPLES}
\label{results}

\begin{figure}
\includegraphics[scale=0.8]{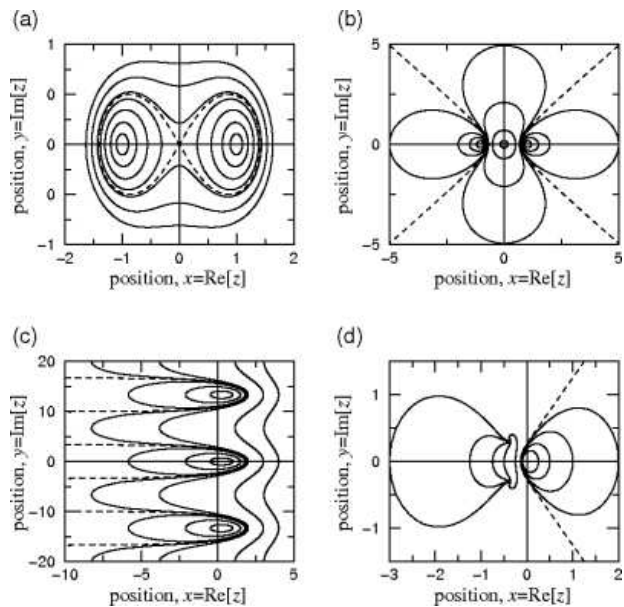}
        \caption{Complex quantum trajectories for a representative sampling of
                 one-dimensional nondegenerate stationary states for Hamiltonians
                 of the \eq{hamiltonian} form. All units are atomic units.
                 Dashed lines indicated separatrix
                 trajectories, leading to branch cuts in the map $f^{T/2}(z)$.
                 See text for additional discussion.
                 Specific states are as follows: (a) System II, Harmonic Oscillator
                 First Excited State; (b) System III, Symmetric Double-Peaked
                 Ground State; (c) System IV, Morse Oscillator Ground State;
                 (d) System V, Asymmetric Single-Peaked Ground State.}
        \label{trajfig}
\end{figure}

To demonstrate the range of dynamical behaviors available, both for
analytic and non-analytic $v(z)$, we consider a representative
sampling of $d=1$ systems, some of which have been considered in the
previous literature \cite{leacock83,john02,yang05a,yang06b,chou07}.
In each case, the Hamiltonian is of the form
\eb
     \H = -{\hbar^2 \over 2m} \prtsq{z} + V(z), \label{hamiltonian}
\ee
with $m=\hbar=1$. All units may therefore be taken to be atomic units.
Complex quantum trajectories for systems II--V are
presented in Fig.~\ref{trajfig}. In every case, \eq{newweak} has been
confirmed numerically, for those trajectories that recross the real axis.
The velocity field $v(z)$ is obtained using \eq{vee}; stationary points
are obtained by solving $v(z_0)=0$; corresponding trajectory periods are
obtained as per the end of Sec.~\ref{analytic}.

\subsection{System I: Harmonic Oscillator Ground State}
\ea{
     \text{potential:}        & \qquad &  V(z) = z^2/2 \nonumber \\
     \text{wavefunction:}     & \qquad &  \psi(z) = \exp(-z^2/2)
                                                             \nonumber\\
     \text{velocity:}         & \qquad &  v(z) = i z \nonumber \\
     \text{stationary point:} & \qquad &  z_0 = 0 \nonumber \\
     \text{period:}           & \qquad &  T = 2 \pi \nonumber}
This system conforms to the special symmetric case, discussed
after \eq{newweak}, for which $|\rho'(x_i)|=|\rho'(x_f)|$. Due
to symmetry, $x_f = -x_i$ corresponds to the first real-axis crossing at
time $T/2$. Consequently, $dx_f = -dx_i$, $\rho(x_i) = \rho(x_f)$,
and weak probability conservation is satisfied.
This is the {\em only} example considered for which
\eq{pconweak} is correct. Also, rather uncharacteristically,
$f^{\Delta_t}(z) = e^{i \Delta_t} z$ is globally analytic. The
complex trajectory orbits are concentric circles centered at the
origin that move counterclockwise.

\subsection{System II: Harmonic Oscillator First Excited State}
\ea{
     \text{potential:}        & \qquad &  V(z) = z^2/2 \nonumber\\
     \text{wavefunction:}     & \qquad &  \psi(z) = z \exp(-z^2/2)
                                                            \nonumber \\
     \text{velocity:}         & \qquad &  v(z) = i (z^2-1)/z \nonumber \\
     \text{node:}             & \qquad &  z = 0 \nonumber \\
     \text{stationary point:} & \qquad &  z_0 = \pm 1 \nonumber \\
     \text{period:}           & \qquad &  T = \pi \nonumber }
The first excited state of the harmonic oscillator has a node at the origin.
As per the discussion at the start of Sec.~\ref{analytic}, this implies that
the velocity field $v(z)$ is meromorphic, with a simple pole at the
$z=0$ node. Through this node, separatrices partition two sets of closed
trajectories (period $\pi$) around the two stationary points. Note that since
these stationary points lie on the real axis, the nearby trajectories
must be concentric circles ($A_1$ is pure imaginary). There are also
larger trajectory orbits, also closed (period $2\pi$), that surround both
stationary points. Thus, trajectories on either side of the separatrices
get mapped to very different locations, so for this example,
$f^{T/2}(z)$ is not globally analytic, or even meromorphic.

\subsection{System III: Symmetric Double-Peaked Ground State}
\ea{
     \text{wavefunction:}     & \qquad &  \psi(z) = \exp(-z^4/2 + z^2)
                                                    \nonumber  \\
     \text{velocity:}         & \qquad &  v(z) = 2 i z(z^2-1)\nonumber \\
     \text{stationary point:} & \qquad &  z_0 = 0,\pm 1 \nonumber \\
     \text{period:}           & \qquad &  T = \pi, \pi/2 \nonumber }
There are no wavefunction nodes, and therefore no simple poles in
the velocity field $v(z)$, nor in the trajectory map $f^{T/2}(z)$.
Although $v(z)$ is globally analytic, $f^{T/2}(z)$ is
not. All trajectories except for the separatrices are closed, each
surrounding exactly one of the three stationary points. This is the
first case considered for which there are separatrix trajectories
when $v(z)$ itself is globally analytic.

\subsection{System IV: Morse Oscillator Ground State}
\ea{
     \text{potential:}        & \qquad &  V(z) = \exp(-2\sqrt{2}z/3)\nonumber\\
                              & \qquad & - 2\exp(-\sqrt{2}z/3)  \nonumber \\
     \text{wavefunction:}     & \qquad &  \psi(z) = 6\,18^{1/4}\nonumber\\
                              & \qquad & \times\exp\!\of{-3 e^{-\sqrt{2} z/3} - 5\sqrt{2} z /6} \nonumber \\
     \text{velocity:}         & \qquad &  v(z) = i (5-
                     6 e^{-\sqrt{2}z/3})/3\sqrt{2} \nonumber \\
     \text{stationary point:} & \qquad &  z_0 = 3 \log(6/5)/\sqrt{2} + i k 3 \sqrt{2} \pi\nonumber\\
                              & \qquad & \qquad\text{with $k$ an integer} \nonumber \\
     \text{period:}           & \qquad &  T = 18\pi/5 \nonumber }
The velocity field $v(z)$ is globally analytic, but $f^{T/2}(z)$ is
not. Also, $v(z)$ is periodic in the imaginary direction, over a
distance $3 \sqrt{2} \pi$. Thus, the stationary point along the real
axis, and surrounding ``librational'' closed trajectories, are
duplicated at regular intervals away from the real axis. In
addition, there is a family of open, ``hindered rotational''
trajectories, on the right side of the figure, that traverse all
unit cells.

\subsection{System V: Asymmetric Single-Peaked Ground State}
\ea{
     \text{wavefunction:}     & \qquad &  \psi(z) = \exp(-z^4 - z^2/2 - z^3)
                                                               \nonumber \\
     \text{velocity:}         & \qquad &  v(z) = i (z + 3z^2 + 4z^3)
                                                                \nonumber \\
     \text{stationary point:} & \qquad &  z_0 = 0, -3/8 \pm i \sqrt{7}/8
                                                                \nonumber \\
     \text{period:}           & \qquad &  T = 2\pi \nonumber }
The velocity field $v(z)$ is globally analytic, but $f^{T/2}(z)$ is
not. The stationary point at $z_0=0$ is surrounded by closed
trajectories with period $2 \pi$, on the right side of the figure.
In addition, there is a pair of complex-conjugate stationary points
off of the real axis, such that the one in the upper half plane is a
repeller, and the other an attractor. On the left side of the figure
is a family of aperiodic, doubly spiraling trajectories that connect
the repeller to the attractor. These trajectories cross the real
axis only once.

\section{CONCLUSIONS}
\label{conclusions}

We conclude with a brief summary of what has been achieved here.
First, it seems evident that any analysis of quantum probability
flux on complexified space requires generalized complex conjugation
of the form of \eq{conj}. This results in analyticity of the
requisite quantities such as probability density---but much more
importantly, leads to complexified flux relations [e.g.,
\eq{compcont}] that are physically relevant because they do not
depend explicitly on the potential energy. On the other hand, the
most straightforward complex generalization of the flux continuity
relation to emerge from this work, i.e. \eq{compcont}, corresponds
to the $\RE{v}(z,t)$ rather than the $v(z,t)$ velocity field. Even
relaxing to the weaker condition of \eq{pconweak} is insufficient to
achieve probability conservation for $v(z,t)$. Yet even if this
condition were somehow satisfied, it might not be very useful for
most multidimensional applications, because few if any trajectories
recross all of the real coordinate axes simultaneously (even for
separable systems, if the frequencies are incommensurate).

The above suggests that a different choice of complex velocity field
might be more advantageous, though it is not clear at present how to
construct such a field. One possible approach, inspired by coherent
state initial value representations \cite{tannor}, might be to
regard $z$ and $z^*$ as completely independent quantities, e.g. in
formulating partial derivative expressions for the flux and its
divergence. This might in principle require a pair of distinct
complex trajectories, one on $z$-space and another on $z^*$-space,
which somehow get combined together to form the time-evolving
probability density, $\rho[z(t),z^*(t),t]$. Perhaps even two time
coordinates would be involved.

On the other hand, there is still some hope for the $v(z,t)$ approach,
as well. In particular, as per the discussion following \eq{plug}, the
fact that real $dx_i$ implies real $dx_f$ under $v(z,t)$ dynamics might
turn out to be quite useful. Also, in certain special cases, it may turn
out that \eq{pconweak} is {\em approximately} satisfied---well enough to
enable calculations for large systems. In any event, it is hoped that
this initial foray may enable subsequent developments towards achieving
probability conservation for complex quantum trajectories, or at least,
provides a useful framework for analysis of these important issues.

\begin{acknowledgments}
This work was supported by a grant from
The Welch Foundation (D-1523), and by a Small Grant for Exploratory Research
from the National Science Foundation (CHE-0741321).
The author wishes to express gratitude to Yair Goldfarb, Jeremy Schiff,
David Tannor, and Robert Wyatt, for interesting discussions on complex
Bohmian mechanics. Toufik Djama is also acknowledged, for useful discussions
leading to \eq{rsol}.
Jason McAfee is also acknowledged for his aid in converting this manuscript to an electronic format suitable for the arXiv preprint server.
\end{acknowledgments}

%
%

%


\end{document}